# High-Performances AlGaN-based DUV-LED via Under-Level Multiple Quantum Well Configuration

Mohammad Amirul Hairol Aman[1,2], Nurul Fathinah Azrisham[1,2], Ahmad Fakhrurrazi Ahmad Noorden[*,1,2], Wan Hazman Danial[3], Muhamad Zamzuri Abdul Kadir[1,2]

[1] *Centre for Advanced Optoelectronics Research (CAPTOR), Department of Physics, Kulliyyah of Science, International Islamic University Malaysia. 25200 Kuantan, Pahang, Malaysia.*
[2] *IIUM Photonics and Quantum Centre (IPQC), Kulliyyah of Science, International Islamic University Malaysia, 25200, Kuantan, Pahang, Malaysia.*
[3] *Department of Chemistry, Kulliyyah of Science, International Islamic University Malaysia, 25200 Kuantan, Pahang, Malaysia.*

Corresponding Author [*]: *fakhrurrazi@iium.edu.my*

**Abstract**:

Low internal and external quantum efficiencies in high Aluminium content AlGaN-based deep-ultraviolet light-emitting diode (DUV-LED) occurred due to strong polarization effects, spontaneous and piezoelectric polarization, at the interface between two materials. It also leads to a low carrier confinement and Quantum Confined Stark Effect (QCSE), contributing to the efficiency droop of the DUV-LED. This work demonstrates an under-level MQW configuration implemented in a DUV-LED with a 257 nm emission wavelength. Three DUV-LED structures, above-, same- and under-level MQW were investigated, covering important optoelectronics properties such as energy band diagram, carrier concentrations, radiative recombination rates and electric field distribution. It is found that the quantum efficiencies, luminescence intensity and light output power of the under-level configuration has been enhanced by nine-, ten- and five-folds, relative to the above-level MQW configuration.

# 1.0 INTRODUCTION

Deep ultraviolet (DUV) light-emitting diodes (LEDs), emitting light in the wavelength range of 200-280 nm, represent a promising technology with diverse applications in fields such as water [1,2] and air purification [3], medical sterilization [4], and environmental sensing [5,6]. While DUV-LEDs offer significant advantages over traditional light sources, including energy efficiency [1,2], long lifespan [7], and environmental friendliness [2,4], their widespread adoption is hindered by several challenges, primarily related to their low efficiency.

The state-of-the-art DUV-LEDs typically exhibit an external quantum efficiency (EQE) below 10% for wavelengths between 250 nm and 260 nm [8–10]. EQE, a crucial performance metric, is the product of internal quantum efficiency (IQE) and light extraction efficiency (LEE) [11–13]. IQE, representing the efficiency of radiative recombination, is significantly limited in DUV LEDs. The low IQE of DUV LEDs is primarily attributed to three factors: carrier overflow [14], inefficient carrier injection into the active region [12], and polarization effects [15]. Carrier overflow results in non-radiative recombination at high injection currents due to a substantial portion of carriers (electrons and holes) fail to recombine radiatively, while inefficient carrier injection limits the number of photons generated in the active region. This limitation arises from the high potential barriers at the interfaces between different semiconductor layers. Polarization effects create a built-in electric field in DUV LEDs which leads to significant Quantum-Confined Stark Effect (QCSE). This effect separates electron-hole pairs, reducing the probability of radiative recombination and further degrading the device performance. Addressing these fundamental challenges is crucial for improving the efficiency of DUV-LED and unlocking their full potential for a wide range of applications.

Among the famous strategy to reduce the effect of efficiency droop is the implementation of the electron blocking layer (EBL) and hole blocking layer (HBL). The EBL aims to reduce the overflow of the electron from the active region into the p-region, while simultaneously, improving the injection of hole from the p-region into the active region [16]. Numerous EBL designs have been proposed and investigated, such as the superlattice step-doped EBL (SDSL-EBL) [17], polarized electric field reservoir EBL (PEFR-EBL) [18], superlattice EBL (S-EBL) [19]. Contrarily, the purpose of the HBL is to block the overflow of hole from the active region into the n-region and also improve the injection of electron from n-region into the active region [20]. Normally, the HBL is paired with EBL, aiming to increase the confinement of carrier by mitigating the overflow effects, and improving the injection of carrier.

Wang et al proposed m-shaped HBL and w-shaped EBL in AlGaN-based DUV-LED to improve the light output power (LOP), emission spectra intensity and IQE [21]. The combination of m-shaped HBL and w-shaped EBL effectively improve the energy barrier height leading to higher carrier confinement. The probability of radiative recombination process to occur in the quantum well is further enhanced by the overlapping of carrier wavefunction, originating from the reduction of polarization effects in the active region [21]. Another work on HBL-EBL in AlGaN-based DUV-LED is by Shi et al, which investigates an irregular sawtooth HBL and EBL [22]. The LOP and IQE of the proposed configuration enhanced significantly relative to the reference structure, showcasing the effectiveness of the irregular sawtooth EBL and HBL in fulfilling their roles [22]. The efficiency droop of the device also has been toned down greatly, indicating that the carrier leakage is significantly reduced. The irregular sawtooth HBL and EBL also managed to optimize the effective carrier barrier height, leading to higher carrier confinement within the active region.

Another approach in optimizing the performance of the device is through band-engineering of multiple quantum well (MQW) and barrier (MQB). Lang et al proposed an asymmetrical concave quantum barriers (AC-QB) to modulate the velocity of the carrier [23]. The enhancement in performances is shown in the increment of electroluminescence intensity, EQE and LOP. The hole concentrations within the MQW, obtained via AC-QB implementation, improved by a large margin compared to the conventional to bulk MQB, leading to increase in radiative recombination rates and better performances [23]. Yu et al proposed an innovative design, Al-composition graded quantum barrier (Al-GQB) where the aluminium composition of the QB gradually increased from the first QB to the last QB [24]. The Al-GQB aims to suppress the electron leakage and increase hole injection efficiency of the AlGaN-based DUV-LED. The LOP and EQE of the device improved significantly, attributed to the optimized effective carrier barrier height of the MQB and the increased in overlapping carrier wavefunction [24]. The strategies of band-engineering also were applied in the MQW such as the unidirectional graded QW [25], AlGaN-delta-GaN QW [26] and graded composition MQW [27].

Xing et al proposed an interesting approach to improve DUV-LED performance which is via dual polarization carrier injection layer [28]. The strategy is to increase the material composition of the electron injection layer, n-AlGaN gradually until it reaches the composition of QB and gradually decrease the material composition of the hole injection layer, p-AlGaN down to p-GaN. This will result in reduced polarization effects between the layers, leading to an optimized effective carrier barrier height. The LOP and EQE of the device with dual polarization carrier injection layer alleviated significantly, relative to the reference structure.

Efficiency droop, originating from the carrier overflow and low carrier injection is a major obstacle in the progress of DUV-LED technology [29,30]. The efficiency of the device also aggravates further due to the high polarization effects, spontaneous and piezoelectric polarizations, which leads to separation of carrier wavefunction in the MQW and degrades the radiative recombination rates [31]. Not only that, but high precision technology is also required to fabricate the proposed complicated structures. Inaccuracy in tuning the composition of the material or other important parameters will degrade the optoelectronics properties, deteriorating the overall performances of the DUV-LED [32]. Significant lattice mismatch of epitaxy layers growth also would lead to threading dislocations density, intensifying non-radiative recombination centre and lowering the device performance [33].

In this work, we aim to overcome the issue of efficiency droop by mitigating the overflow of carrier, simultaneously, improving the carrier injection while maintaining unsophisticated configuration in order to ease the fabrication process of the device. We proposed an under-level MQW configuration which is much easier to realize, for AlGaN-based DUV-LED with ~260 nm emission wavelength. Two other LEDs, above-level and same-level MQW, which is a variation of the proposed configurations also is simulated to comprehensively understand the physics and scientific reasoning behind the phenomenon occurred. Key performance parameters such as the IQE, EQE, LOP and luminescence spectrum are investigated. The factors leading to the performances of the device also is included in the analysis, covering important optoelectronics properties, such as energy band diagram, carrier concentration distribution, radiative recombination rates and strength of electric field.

## 2.0 MODEL-THEORY & DEVICE PARAMETERS

The simulation of the AlGaN-based DUV-LED is performed by using One Dimensional Poisson, Drift-Diffusion and Schrödinger Solver (1D-DDCC) developed by Yuh-Renn Wu [34–36]. The software utilizes a drift-diffusion model (DDM) to compute the transportation of the carrier within semiconductor materials, while considering the quantum physics effects. The DDM consists of the three main equations, current density equation, continuity equation and Poisson equation, that are solved self-consistently [37]. The current density equations for electron, $J_n$ and hole, $J_p$ can be expressed as

$$J_n = q\mu_n n E + q D_n \nabla n \qquad (1)$$

$$J_p = q\mu_p p E + q D_p \nabla p \qquad (2)$$

where $\mu$, $D$, $E$ and $q$ refer to the carrier mobility, carrier diffusion coefficient, electric field and elementary charges. The term $n$ and $p$ represent the electron and hole concentrations, respectively, which can be obtained by solving the Fermi-Dirac distribution and density of states associated with the Schrödinger equations. The continuity equation for electron and hole are given by

$$\frac{\delta \Delta n}{\delta t} = G_{ext} - \frac{\Delta n}{\tau_n} - \mu_n \frac{\delta E}{\delta x} \Delta n + \mu_n E \frac{\delta \Delta n}{\delta x} + D_n \frac{\delta^2 \Delta n}{\delta x^2} \qquad (3)$$

$$\frac{\delta \Delta p}{\delta t} = G_{ext} - \frac{\Delta p}{\tau_p} - \mu_p \frac{\delta E}{\delta x} \Delta p - \mu_p E \frac{\delta \Delta p}{\delta x} + D_p \frac{\delta^2 \Delta p}{\delta x^2} \qquad (4)$$

where $G_{ext}$ is the generation due to external factors and $\tau$ is the time taken for scattering events of the carrier to occur. The Poisson equation is determined as

$$\frac{dE(x)}{dx} = \frac{q(p(x) - n(x) + N_D(x) - N_A(x))}{\varepsilon} \qquad (5)$$

and $\varepsilon$ is the dielectric permittivity of the material. The carrier mobility which depends on the doping concentration, $N$ in the AlGaN-based semiconductor is obtained via Caughey-Thomas Model and is written as [17]

$$\mu(N) = \mu_{min} + \frac{\mu_{max} - \mu_{min}}{1 + \left(\frac{N}{N_{ref}}\right)^{\alpha}} \quad (6)$$

The $\mu_{min}$, $\mu_{max}$, $N_{ref}$ and $\alpha$ is parameters obtained by fitting with the experimental results for each electron and hole. The $\mu_{min}$, $\mu_{max}$, $N_{ref}$ and $\alpha$ for electron are 132 cm²V⁻¹s⁻¹, 302 cm²V⁻¹s⁻¹, $1.0 \times 10^{17}$ cm⁻³ and 0.29, respectively [17]. The parameters $\mu_{min}$, $\mu_{max}$, $N_{ref}$ and $\alpha$ for hole are 2 cm²V⁻¹s⁻¹, 10 cm²V⁻¹s⁻¹, $1.0 \times 10^{17}$ cm⁻³ and 0.395, respectively [17]. The EQE of the LED is computed through [38]

$$EQE = IQE \times LEE \quad (7)$$

where LEE and IQE refer to the light extraction efficiency and internal quantum efficiency. The IQE can be obtained through the ABC model which is given by [39]

$$IQE = \frac{Bn^2}{An + Bn^2 + Cn^3} \quad (8)$$

where $n$ is the carrier density. The Shockley-Read-Hall (SRH), bimolecular radiative and Auger recombination coefficient are represented by $A$, $B$ and $C$ [39].

The AlGaN-based DUV-LED is grown on a AlN substrate, followed by AlGaN buffer layers and 3 µm n-AlGaN electron injection layer (EIL) doped with $5 \times 10^{18}$ cm⁻³ Silicon (Si). Then, an active region consists of multiple quantum well (MQW) alternate with multiple quantum barrier (MQB) grown on the electron injection layer. The QW and QB are made up of 4 nm undoped $Al_{0.60}Ga_{0.40}N$ and 12 nm undoped $Al_{0.70}Ga_{0.30}N$. Immediately after the active region is a 20 nm p-$Al_{0.85}Ga_{0.15}N$ electron blocking layer (EBL) dope with $5 \times 10^{17}$ cm⁻³ Magnesium (Mg). The hole injection layer (HIL) consists of 15 nm p-AlGaN doped with $1 \times 10^{19}$ cm⁻³ Mg. The DUV-LED is capped with 35 nm p-GaN and $1 \times 10^{20}$ cm⁻³ Mg doped. The ionization energy of the Mg is set to increase with aluminium components, from 170 meV to 470 meV [16]. The SRH lifetime, bimolecular radiative and Auger recombination coefficients are set to 10 ns, $2 \times 10^{-11}$ cm³s⁻¹ and $2 \times 10^{-31}$ cm⁶s⁻¹, respectively. The degree of polarization, which occurred due to spontaneous and piezoelectric polarizations, is considered to be 50% of the theoretical value [40] and the band off-set is set to 0.63/0.37. The LEE to compute the EQE is assumed to be 10% [41].

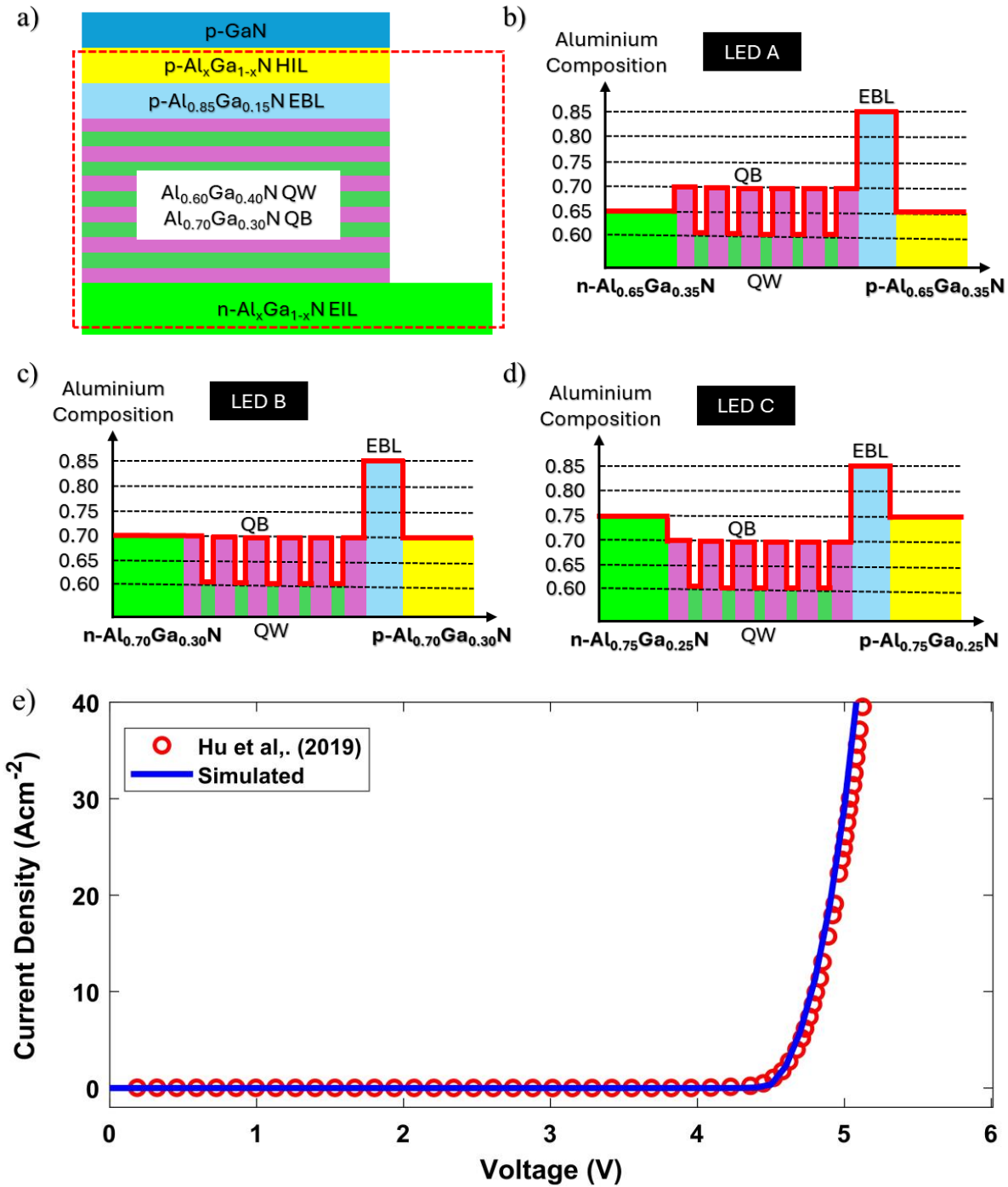

Figure 1. a) The schematic diagram of the DUV-LED. The aluminium composition, starting from electron injection layer (EIL), n-AlGaN to the hole injection layer (HIL), p-AlGaN, of the above-level, same-level and under-level MQW is denoted as b) LED A, c) LED B and d) LED C, respectively. The e) Validation of DUV-LED via IV characteristic with research work by [42].

The simulation software used in this research is 1D-DDCC, developed by Prof. Yuh-Renn Wu. The validation of the software was conducted by comparing the results generated from the 1D-DDCC simulations with experimental data from a fabricated DUV-LED structure reported by Hu et al [42]. The device parameters of the simulated DUV-LED can be found in [42,43]. Figure 1e) presents the IV characteristics obtained from the 1D-DDCC simulations, which demonstrate strong agreement with the IV characteristics of the fabricated DUV-LED. This validation confirms the reliability of the 1D-DDCC software, making it a suitable tool for further research in DUV-LED design and optimization.

**4.0 RESULT & DISCUSSION**

Three DUV-LEDs with different configurations were simulated. The optoelectronics properties of the structures such as energy band diagram, carrier concentrations, radiative recombination rate, electron leakage and electric field distribution were discussed. The performances of the DUV-LED also were analyzed to determine the benefits of the proposed structure. Figure 2 shows the IQE, EQE, luminescence intensity and LOP of the three DUV-LED.

Based on Figure 2a) and b), the LED C has the highest peak IQE and EQE followed by LED B and LED A. The efficiency droop of the LEDs also was measured at 300 $Acm^{-2}$ as shown in Figure 2a), where LED A experiences the lowest droop followed by LED B and LED C. The peak IQE, EQE, and efficiency droop of the LEDs is tabulated in Table 1. It is noteworthy that although LED C experiences higher efficiency droop compared to LED A, the efficiency for all current injections of LED C is significantly higher than LED A. This differences occurred due to the difference in the confinement ability of the MQW-MQB regions, and the severity of the electron overflow and low hole injection.

Table 1. Summary of key performance parameters for LED A, LED B and LED C.

|  | Peak IQE (%) | Peak EQE (%) | Efficiency Droop at 300 $Acm^{-2}$ (%) | Lum. Int. ($1 \times 10^{20}$ a.u.) | Wavelength (nm) | LOP at 300 $Acm^{-2}$ (mW) |
|---|---|---|---|---|---|---|
| LED A | 5.48522 | 0.5485 | 33 | 0.56711 | 261.445 | 7.64408 |
| LED B | 32.2886 | 3.2288 | 48 | 5.69919 | 257.971 | 24.5766 |
| LED C | 50.3004 | 5.0300 | 41 | 9.51982 | 257.757 | 41.9246 |

Additionally, LED C shows the highest luminescence intensity relative to LED A and LED B as shown in Figure 2c). The huge discrepancy between LED C and LED A is nearly ten-fold, whereas LED B is two-fold, highlighting the high carrier confinement within the

MQW and high photon emission. Furthermore, the emission wavelength of LED A, LED B and LED C is 257.757 nm, 257.971 nm and 261.445 nm. However, LED A experiences high parasitic emissions which occur at wavelength 258.562 nm. Parasitic emissions occurred due to two reasons which are band-to-band recombination or recombination through defects [44], leading to a different energy bandgap, subsequently different wavelength emissions. The luminescence intensity and emission wavelength of the three LEDs is tabulated in Table 1. The LOP of the LEDs also were measured at 300 Acm$^{-2}$ as observed in Figure 3d), showing that LED C has the highest value, followed by LED B and LED C. This is also attributed to the high carrier confinement obtained by LED C, improving the LOP by two-fold of LED B and four-fold of LED A. The LOP of the three LEDs also is tabulated in Table 1. To understand the performances achieved by these LEDs, the energy band diagrams, and corresponding carrier concentrations are analyzed. Figures 3, 4 and 5 depict the energy band diagrams, electron and hole concentrations for each LEDs.

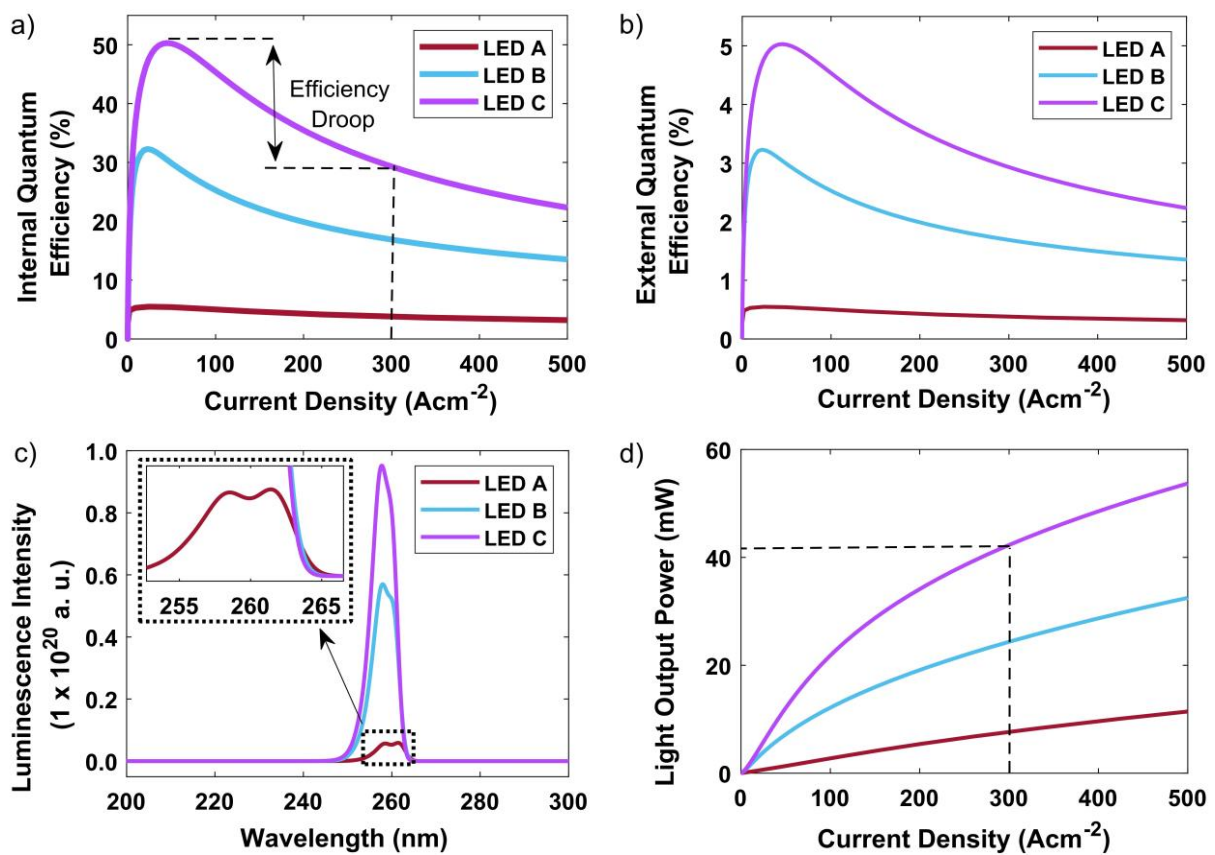

Figure 2. The a) internal quantum efficiency, b) external quantum efficiency, c) luminescence spectrum and d) light output power of LED A, B and C, respectively. The in-set in c) shows the parasitic emission of LED A.

Based on Figure 3a), the electrons are injected into the MQW region through the electron injection layer, penetrating the first QB layer and moving into the MQW region. However, in between the electron injection layer, n-$Al_{0.65}Ga_{0.35}N$ and the first QB, $Al_{0.70}Ga_{0.30}N$, a potential energy barrier, $\Phi_{e,1}$ is formed. Notice that the aluminium composition of electron injection layer is lower compared to the first QB, inducing the polarization effects, creating a potential energy barrier through band-bending and leading to the electrons requiring more energy to cross the barrier via thermionic emission [45]. The formation of potential energy barrier at that position is due to the lattice mismatch between electron injection layer and the first QB, altering the atomic distance and increasing the energy required to enter the MQW. In conjunction with the formation of potential energy barrier in the conduction band, a quantum well also unintentionally developed in the valance band, trapping the hole and may contribute as a parasitic hole reservoir [16], degrading the optoelectronics performances. This phenomenon would lead to the reduction of electron injections into the MQW, leading to lower electron confinement.

The potential energy barrier of the EBL for LED A also is measured where the effective electron potential energy barrier height, $\Phi_{e,LED\ A}$ is 660.3 meV and the effective hole potential energy barrier height, $\Phi_{h,LED\ A}$ is 1034.4 meV. The EBL main purposes are to block the overflow of the electron from the MQW region into the p-region and increase the injection of hole from the p-region into the MQW region. These would lead to increment in both carriers, enhancing the confinement and improving the optoelectronics properties of the LED. The conduction band of LED A also exhibits an "inverted bowl" shape due to the formation of the potential energy barrier at approximately 5500 nm and the EBL, leading to the difficulty in the injection of electron into the MQW region.

Figure 3b) shows the electron distribution over the MQW and EBL regions for LED A. The parasitic electron reservoir occurred at 5613 nm is due to the polarization effects caused by the lattice mismatch between the EBL, p-$Al_{0.85}Ga_{0.15}N$ and the hole injection layer, $Al_{0.65}Ga_{0.35}N$, reducing the electron confinement in the MQW region. Moreover, the electron concentration gradually decreased from the first QW to the fifth QW. This is because the electrons require more energy as they penetrate the active region. Hence, the formation of the potential energy barrier, $\Phi_{e,1}$ reduces the energy of the electron and becomes more challenging for them to cross the QBs. Figure 4c) depicts the hole concentration for MQW and EBL regions. Similarly, the parasitic hole reservoir also is formed at the EBL interfaces, located at 5592 nm, between the last QB, $Al_{0.70}Ga_{0.30}N$ and the EBL, p-$Al_{0.85}Ga_{0.15}N$. This is caused by the polarization effects, originating from the lattice mismatch between the two layers. It also will reduce the hole confinement in the active region, degrading the performance of the LED. It is noteworthy to mention that there is accumulation of holes at 5500 nm as observed in Figure 3c). This is the effect of the potential energy barrier formation, creating an unwanted QW, allowing the holes to trap in that location, contributing to less hole confinement in the MQW region.

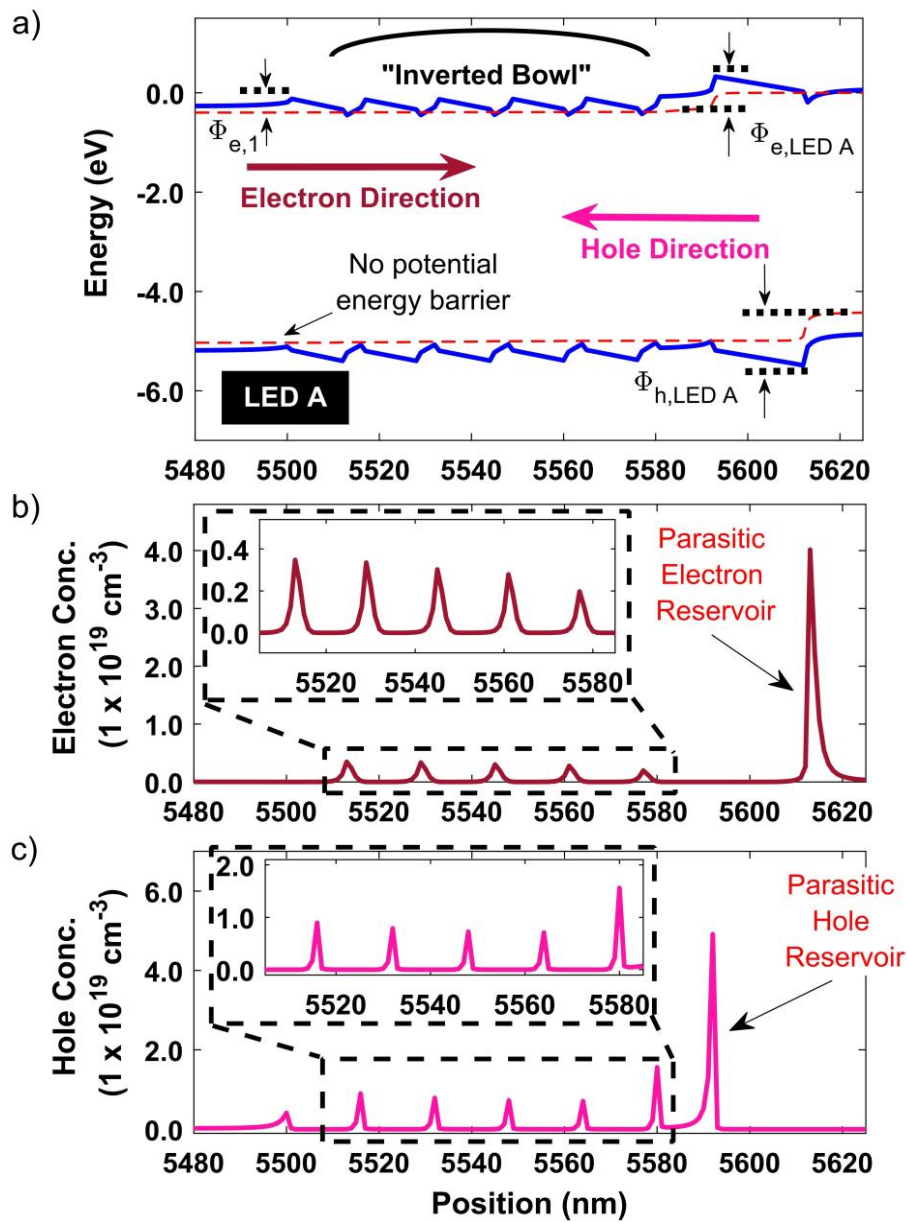

Figure 3. The a) energy band diagram, b) electron concentration and c) hole concentration of LED A for MQW and EBL region. The in-let in b) and c) show the carrier concentration in the MQW region.

Figure 4 shows the energy band diagram and carrier concentration of LED B. Based on Figure 4a), there is no potential energy barrier formed between the electron injection layer, n-$Al_{0.70}Ga_{0.30}N$ and the first QB, $Al_{0.70}Ga_{0.30}N$, since the aluminium composition of these two layers is identical. The lattice-matched at the interface between the two layers does not experience polarizations effects, thus, no band-bending occurred, resulting in no formation of potential energy barrier such as LED A. The absence of the potential energy barrier allows more electron to be injected into the MQW region from the electron injection layer, increasing the confinement of the electron. Note that there is no accumulation of holes, in Figure 4c), located at 5500 nm, compared to LED A. This is due to the lattice-matched layers, mitigating the unwanted QW at that particular location, indicating that the hole is more confined in the MQW region than LED A.

In addition, the potential energy barrier of EBL also is measured, where the potential energy barrier for EBL in the conduction band, $\Phi_{e,LED\ B}$ is 664.0 meV and the valance band, $\Phi_{h,LED\ B}$ is 837.9 meV. The increment in the potential energy barrier height from 660.3 meV of LED A to 664.0 meV of LED B suggests that the blocking ability of EBL is slightly increased, indicating that overflow of electron is reduced. Additionally, the potential energy barrier height for the hole of LED A, 1034.4 meV is decreased to 837.9 meV, facilitating the transportation of hole from the p-region into the MQW region. The conduction band of LED B also has a "flat" shape compared to LED A which has an "inverted bowl" shape. It can be inferred that the "flat" shape conduction band is superior compared to "inverted bowl" shape in contributing to obtain a better optoelectronics performance.

These effects are visible in the electron and hole concentrations of LED B, shown in Figure 4b) and c), compared to LED A. The electron concentration of LED A is in the scale of ~0.4 × $10^{19}$ cm$^{-3}$, whereas LED B is ~1.0 × $10^{19}$ cm$^{-3}$. Furthermore, the parasitic electron reservoir at 5613 nm of LED B, ~3.0 × $10^{19}$ cm$^{-3}$, is lower compared to LED A, ~4.0 × $10^{19}$ cm$^{-3}$, indicating that the electron confinement of LED B is higher than LED A. These enhancements are attributed to the absence of the energy barrier between electron injection layer and first QB, the reduction in parasitic electron reservoir and the improved blocking ability of the EBL. The hole confinement in the MQW region also was enhanced where the average hole concentrations of LED A of all QWs are below 1.0 × $10^{19}$ cm$^{-3}$. Contrarily, the average hole concentrations of LED B for all QWs are above 1.0 × $10^{19}$ cm$^{-3}$ with the fifth QW slightly above the 2.0 × $10^{19}$ cm$^{-3}$ scale. This is credited to the decrease in potential energy barrier height of the hole, leading to an increase in hole confinement in the MQW region. On the other hand, there is no visible change in the parasitic hole reservoir at 5592 nm between LED A and LED B.

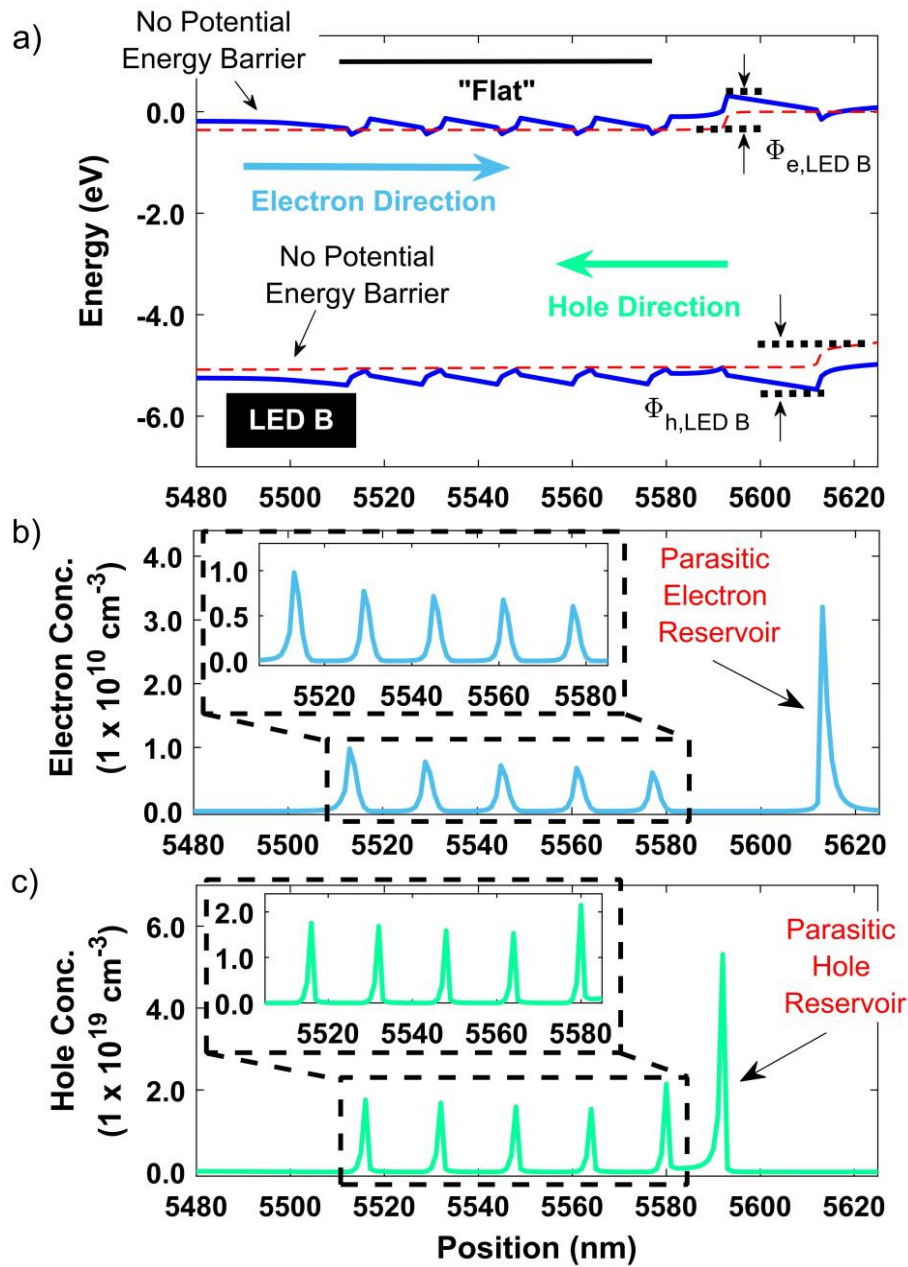

Figure 4. The a) energy band diagram, b) electron concentration and c) hole concentration of LED B for MQW and EBL region. The in-let in b) and c) show the carrier concentration in the MQW region.

Figure 5 depict the energy band diagram and carrier concentration of LED C. In contrast with LED A and LED B, there is a formation of potential energy barrier, $\Phi_{h,1}$ in valance band of LED C at 5500 nm, suggesting that the hole from the MQW region is blocked from overflowing into the n-region. An overflow of hole will result in non-radiative recombination, lowering the hole confinement in the MQW region and degrading the performance of the LED. It is worth mentioning that the formation of potential energy barrier at that particular position is because of the lattice mismatch arising from the difference in aluminium composition of electron injection layer, n-$Al_{0.75}Ga_{0.25}N$ and first QB, $Al_{0.70}Ga_{0.30}N$. The aluminium composition of electron injection layer in LED C is higher compared to the first QB, whereas in LED A, the aluminium composition of the first QB is higher than electron injection layer. This led to different functions of potential energy barrier where in LED A, electron in electron injection layer is blocked from entering the MQW region, implying that the injection of electron is reduced. Contrarily, the function of potential energy barrier in LED C blocks the hole in the MQW region from entering the n-region, resulting in higher hole confinement.

Additionally, the potential energy barrier height of the EBL also is measured where the potential energy barrier height in the conduction band, $\Phi_{e,LED\ C}$ is 651.6 mV and in the valance band, $\Phi_{h,LED\ C}$ in 708.3 meV. The potential energy barrier height of LED C in the conduction band, associated with the transportation of electron, is lower than LED A, 660.3 meV and LED B, 664.0 meV, suggesting that the blocking ability of the EBL is slightly reduced. Furthermore, the potential energy barrier height of hole for LED C is significantly reduced compared to LED A, 1034.4 meV and LED B, 837.9 meV, resulting in increment of hole injection efficiency and improving hole confinement within the MQW region. Although the potential energy barrier height for the electron of LED C is lower than LED A and LED B, it will still confine more electron, since the conduction band of LED C has a "bowl" shape which is suitable for containing the electron in the MQW region. These effects are observed in the carrier concentrations within the MQW and EBL region in Figure 5b) and c).

The electron concentration of LED C is higher than LED B, achieved via the "bowl" shape conduction band having no potential energy barrier blocking the injection of electron into the MQW region and appropriate value for potential energy barrier height of EBL. The parasitic electron reservoir at 5613 nm due to the polarization effects at the EBL interfaces also has been reduced significantly to ~$2.0 \times 10^{19}$ cm$^{-3}$, compared to LED A, ~$4.0 \times 10^{19}$ cm$^{-3}$ and LED B, ~$3.0 \times 10^{19}$ cm$^{-3}$. This implies that the confinement of electron in LED C is higher than LED A and LED B. Furthermore, the parasitic hole reservoir at 5592 nm between the last QB and EBL does not show any visible change compared to LED A and LED B. However, the hole concentration within the MQW region for LED C is significantly higher than LED B, which is the result of potential energy barrier in the valance band acting as a hole blocking layer (HBL). The HBL aim to block the overflow of hole from the active region into the n-region and enhance the injection of electron from n-region into the active region. The presence of the HBL in LED C greatly alleviates the hole concentration in MQW region, leading to improved optoelectronics performance. Figure 7 shows the electron and hole concentration of the three LEDs including their radiative recombination rates in the MQW region.

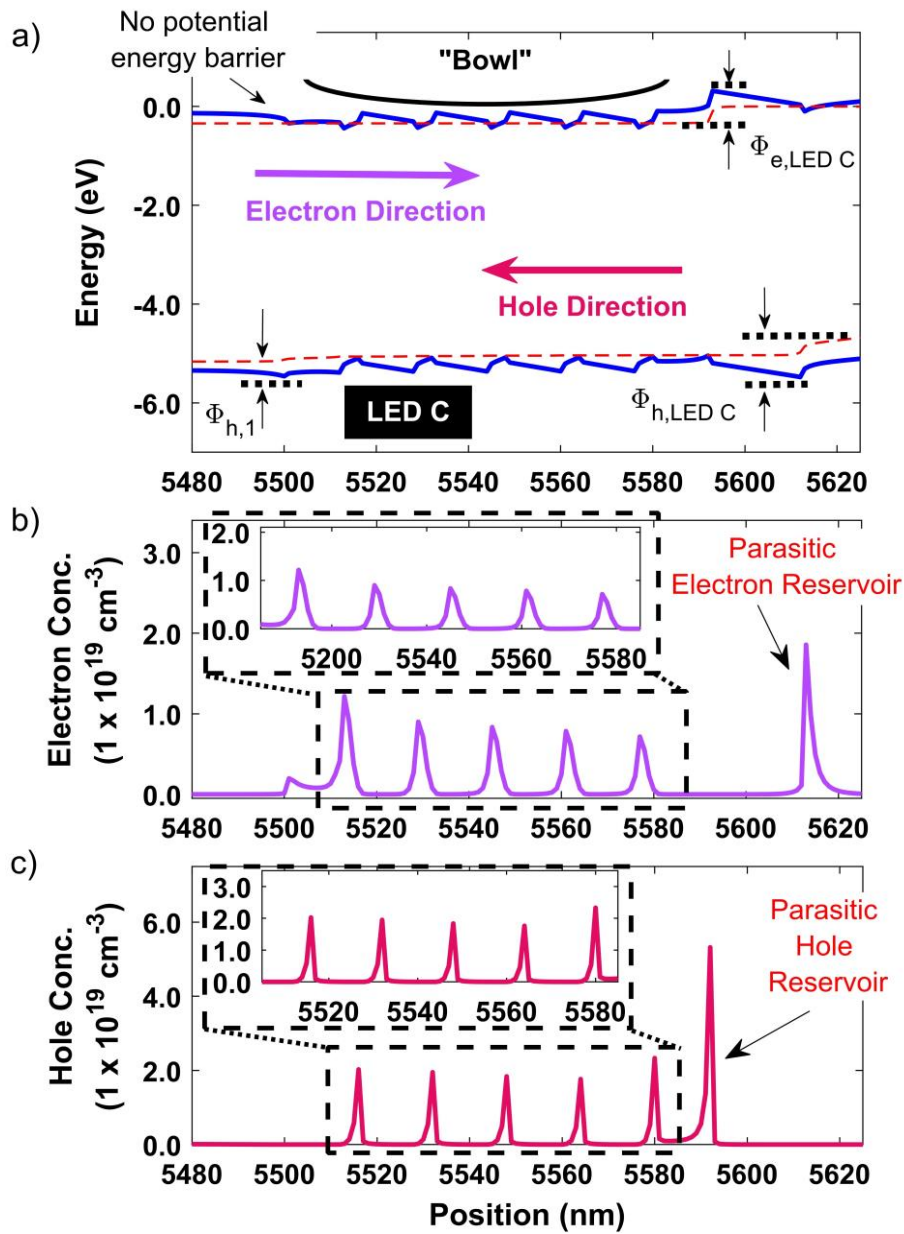

Figure 5. The a) energy band diagram, b) electron concentration and c) hole concentration of LED C for MQW and EBL region. The in-let in b) and c) show the carrier concentration in the MQW region.

Figure 6a) and b) shows the compilation of electron and hole concentrations for the LEDs. The configuration of under-level MQW-MQB in LED C significantly improved the concentration of the carrier within the MQW. This is attributed to the "bowl" shape bands containing the carrier within the MQW through the effect of hole blocking layer attained from the formation of potential energy barrier between the electron injection layer and the first QB. Besides, the addition of the EBL also creates a potential energy barrier, effectively blocking the overflow of electron for leaking into the p-region and degrade the performance of the DUV-LED. The carrier concentration of LED A is the lowest due to the formation of potential energy barrier between the electron injection layer and the first QB, reducing the injection of electron into the MQW region, detrimental to the optoelectronics performances of the LED. Figure 6c) shows the radiative recombination rates for the three LEDs, where LED C has the highest radiative recombination rates in all MQWs, followed by LED B and the lowest, LED A. These results are in-line with the carrier concentrations, as LED C possesses the highest carrier confinement, followed by LED B and LED A. Another factors affecting the radiative recombination rates aside from the confinement of the carriers is the distribution of the electric field in the MQW.

Based on Figure 6d), the magnitude of the electric field in the MQW of LED C is the lowest, followed by LED B and LED A which is the highest. This implies that the polarizations effects at the QW region experience by LED C is the lowest, reducing the band-titling effects, leading to increment in overlapping of carrier wavefunction in the QW. The separation of carrier wavefunction occurred due to polarization effects, titling the bands, making the carrier is concentrated at the band-edge of QW, instead of in the middle of the QW. This phenomenon is known as Quantum Confined Stark Effect (QCSE), which also one of the main factors contributing to the efficiency droop of an LED. Notice that the electric field behaviour at 5500 nm which the formation of the potential energy barrier occurred in LED A and LED C is different. This is explained previously where it occurred due to the aluminium composition differences, leading to lattice mismatch between the layers. Hence, to obtain the hole blocking layer effect in LED C, the aluminium composition of electron injection layer must be higher than the first QB.

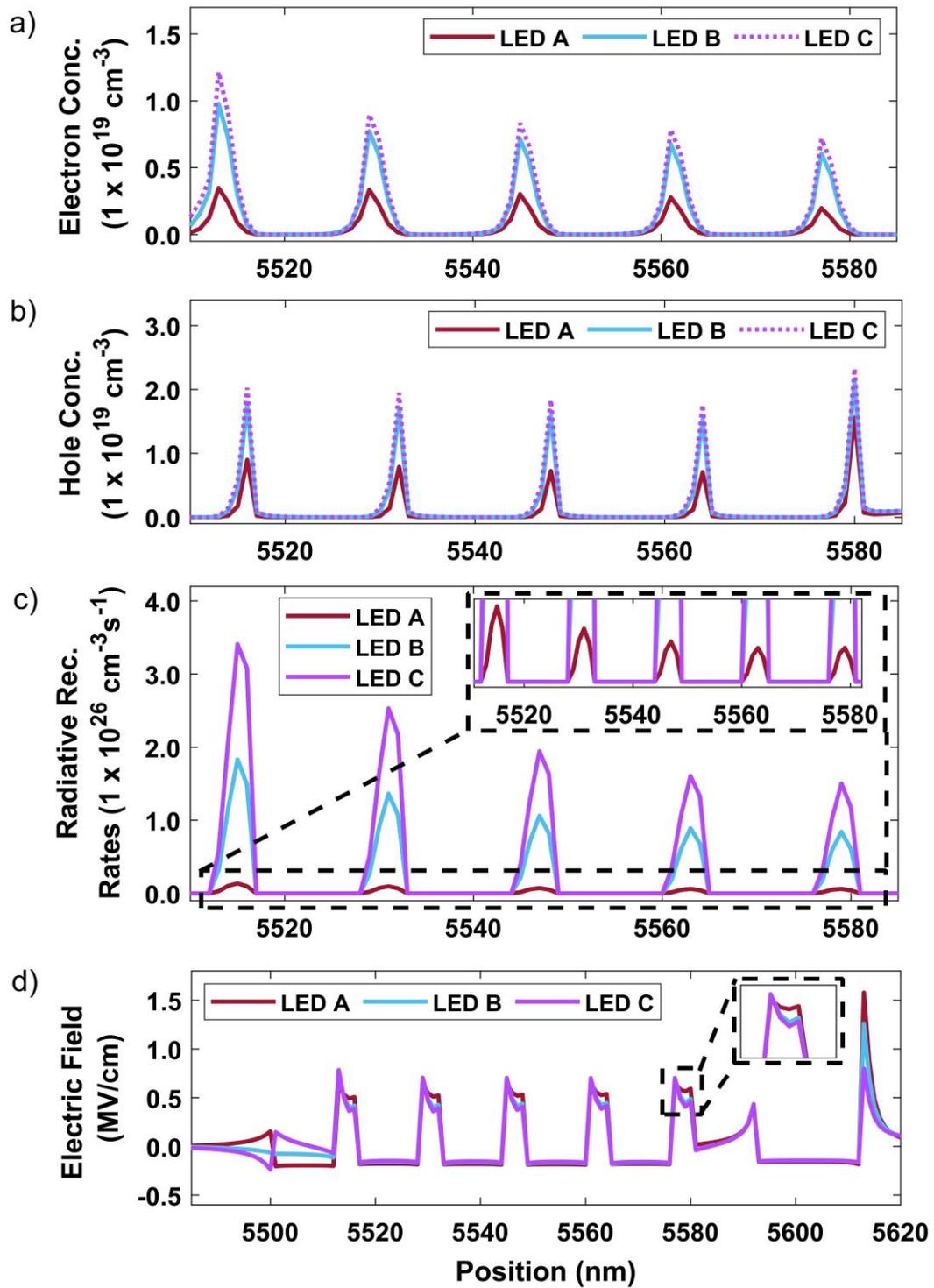

Figure 6. The a) Electron concentrations and b) Hole concentrations of LED A, LED B and LED C. The c) radiative recombination rates and d) electric field of LED A, LED B and LED C, respectively.

## 5.0 Conclusion

In conclusion, the research on the above-, same- and under-level MQW configurations, represented by LED A, LED B and LED C, respectively, have been performed. The performance of the device such as IQE, EQE, luminescence spectrum and LOP has been discussed. The optoelectronics properties of the LED leading to their performances also have been analysed covering several energy band diagram, carrier concentrations, radiative recombination rates, and distribution of the electric field. It is found that the under-level configuration of LED C, leading to a formation of effective barrier height causing the hole blocking effect and improved the electron injection efficiency. Consequently, the electron and hole concentration confined within the MQW increased significantly, enhancing the radiative recombination rates and overall performance of the DUV-LED. It is found that the quantum efficiencies (IQE and EQE), luminescence intensity and LOP of LED C have been enhanced by nine-folds, ten-folds and five-folds, respectively, relative to LED A which has an above-level configuration. The band-engineering of under-level MQW configuration opens a new door in obtaining high-performances DUV-LED with uncomplicated design, reducing the difficulty in the fabrication process.

**Acknowledgements**

The authors would like to acknowledge the support of CAPTOR, IPQC, and the Department of Physics, International Islamic University Malaysia, in terms of facilities and financial by the Ministry of Education (Malaysia) through the Fundamental Research Grant Scheme (Project No.: FRGS 19-033-0641) (References No.: FRGS/1/2018/TK07/UIAM/02/1).

**Conflict of Interest**

The authors declare no conflict of interest.

**Data Availability Statement**

The data that support the findings of this study are available upon reasonable request from the author.